\documentclass[aps,prl,twocolumn,amsmath,amssymb,floatfix]{revtex4}
\usepackage{epsfig,psfrag,subfigure}
\usepackage{graphicx,psfrag,subfigure}
\usepackage{psfrag}

\begin{document}

\title{Aharonov-Bohm interference and fractional statistics in a quantum Hall interferometer}
\author{Eun-Ah Kim}

\affiliation{Stanford Institute for Theoretical Physics and Department of Physics, Stanford, CA 94305}
\date{\today}
\begin{abstract}
We compute the temperature, voltage, and magnetic field dependences of the conductance oscillations of a model interferometer designed to measure the fractional statistics of the quasiparticles in the fractional quantum Hall (FQH) effect.  The geometry is the same as that used in recent experiments~\cite{camino-prl,camino-T,zhou}.  With appropriate assumptions concerning the relative areas of the inner and outer rings of the interferometer, we find the theoretical results, including the existence of {\it super periodic} Aharonov-Bohm (AB) oscillations, to be in remarkably good agreement with experiment.  We then make additional experimental predictions with no adjustable parameters which, if verified, would confirm the proposed interpretation of the experiment as a measurement of fractional statistics.
\end{abstract}
\maketitle

The statistics of identical particles strongly influence their collective quantum behavior in a non-local manner, even in the absence of interactions.
In two spatial dimensions (2D), in addition to Bose and Fermi statistics,  
more exotic fractional~\cite{leinaas77,wilczek82} and even non-abelian braiding statistics are permitted~\cite{pfaffian-short}.  It has long been understood, on theoretical grounds, that the
quasiparticles(qp)/quasiholes(qh) in the FQH phases of the 2D electron gas in a strong perpendicular magnetic field 
obey  fractional braiding statistics, {\it i.e.} they are anyons\cite{laughlin83,halperin84,arovas84}. 

Since the fractional statistics is a tangible implication of topological order that theoretically characterizes different FQH states~\cite{wen95}, 
many interesting proposals for its detection have been put forth~\cite{kivelson90,jain93,chamon-freed-kivelson-sondhi-wen97,safi01,vishveshwara03,kim05,law06}.
Yet Refs.~\cite{camino-T,camino-prl,zhou} were the first to claim to have detected such evidence. 
\textcite{camino-prl} ascribed the observed superperiodic ($\Delta\phi=5\phi_0$) AB oscillation to the fractional statistics obeyed by qp's. 
The four terminal Hall conductance was observed to oscillate around  $\frac{1}{3}\frac{e^2}{h}$ (an indication of 1/3qp's carrying the current ) with period  $5\phi_0$, however a clear theoretical understanding  of this phenomenon has so far been absent. 
This experiment probes a rather subtle situation and  a theoretical model 
which contains essential aspects of the set up is much needed.
In this letter, we study the constructive interference conditions for such a model interferometer and find that $5\phi_0$ oscillation can occur when the topological phase due to {\it both} the fractional statistics and the classical AB effect,  are taken into account. Our calculation of the temperature dependence of the tunneling conductance agrees closely with the data of Ref.~\cite{camino-T} supporting the model  for the observed oscillation.  

FQH liquids are incompressible due to the strong correlations between electrons in a given Landau level.  The different quantum Hall states are characterized, in part, by 
certain rational values of the filling factor $\nu$ (electron density per magnetic flux quantum) at which the system exhibits a quantized Hall conductance, $G_H \!= \!\nu e^2/h$~\cite{tsui82}.  The qp and qh exitations in a given FQH liquid have uniquely determined fractional charge $q^\star$, and 
statistical angle $\theta^\star_\nu$~\cite{arovas84}: the phase change  of the joint wave function for two {\it identical} qp's upon clockwise exchange. For the Laughlin states~\cite{laughlin83} with $\nu=1/(2n+1)$, $q^\star_\nu\!=\!-\nu|e|$ and $\theta^\star_\nu\!=\!\nu\pi$ respectively. 
A qp can be viewed as  a composite object with a charge $q^\star$ bound to a solenoid with flux $\phi_0$ in the direction opposite to that of the external field~\cite{arovas84}. 

For FQH states that are not part of the Laughlin sequence, there are several distinct, but all apparently consistent descriptions~\cite{haldane83,halperin84,jain89,zhang89,lopez91}.  Here we will use Halperin hierarchy, in which the anyonic nature of Laughlin qp's is crucial for the construction of a daughter state~\cite{halperin84}.
Starting with a Laughlin state with filling factor $\nu$, qp charge $q^\star$, and qp statistics $\theta^\star$, a daughter state with filling factor $\tilde{\nu}$ is obtained via condensation of qp of state $\nu$ with
\begin{equation}
\tilde{\nu}\! =\!\nu + \frac{\left({q^\star_\nu}/{e}\right)^2}{\left[2-\left({\theta^\star_\nu}/{\pi}\right)\right]}\!\equiv\!
\nu- q^\star_\nu n_\nu,
\label{eq:hierarchy}
\end{equation}  
where $n_\nu$, the number of Laughlin qp's per area $2\pi l_0^2$ that condense,  is determined by the constraint that many qp wave function obey the fractional statistics determined by $\theta^\star$ (This corresponds to Eqs.(4-5) of Ref.~\cite{halperin84}.). The interferometer of interest involves two distinct FQH states: a Laughlin state $\nu$ and its daughter state $\tilde{\nu}$. Here we focus on the simplest case of $\nu\!=\!1/3$ and $\tilde{\nu}\!=\!2/5$ but the result can be easily generalized.

The model interferometer is shown in the Figure 1. Front gates which confine electrons between two {\it edge states} of opposite chirality,  also define the smooth potential profile that looks like a ``basin'' which can hold more electrons in the central region. In the fractional regime, the basin allows a phase separation between a central puddle of FQH liquid at higher filling $\tilde{\nu}$,   and the surrounding FQH liquid at lower filling $\nu$. 
To derive the interference conditions, we assume the following: {\bf a)} absence of direct tunneling between the outer $1/3$ edge and the inner $2/5$ puddle,
{\bf b)} coherent propagation of  $1/3$ edge qp's which tunnel between  the left moving (upper) edge and the right moving (lower) edge at two point contacts (PC) provided by constrictions, {\bf c)} absence of impurity pinned qp's in the surrounding $1/3$ liquid.
\begin{figure}[t,h,b]
\psfrag{G1}{$\Gamma_1$}\psfrag{G2}{$\Gamma_2$}
\psfrag{nu1}{$\nu$}
\psfrag{nu2}{$\tilde{\nu}$}
\psfrag{B}{$B$}
\psfrag{R}{$R$}
\includegraphics[width=0.3\textwidth]{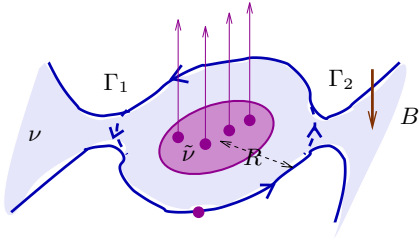}
\caption{The fractional quantum Hall interferometer of our interest. The inner puddle of area $s$, filled with the FQH liquid $\tilde{\nu}\!=\!2/5$, is surrounded by the $\nu\!=\!1/3$ FQH liquid. Quasiparticles of $\nu$ FQH liquid propagating along the outer edge tunnels at two tunneling points.  The coherent tunneling path of $\nu\!=\!1/3$-qp can be approximated by a closed circle of radius $R$ (area $S$) enclosing the flux of $\phi\!=\!-|B|S$.  }
\label{fig:setup}
\end{figure}

We first note that $5\phi_0$ is the unit of flux increment that is associated with an addition of {\it one 1/3 qp} to the puddle, which is the smallest unit of charge that leaves the puddle in its ground state~\cite{jain93}. 
This can be understood from the fact that $n_{1/3}=\frac{1}{5}$ in Eq.~\eqref{eq:hierarchy} and the  total number of $1/3$ qp's in the puddle of area $s$ under magnetic field $B$ is $N\equiv\left\lfloor\frac{n_{1/3}|B|s}{\phi_0}\right\rfloor$ (where $\left\lfloor x\right\rfloor$ reflects the discreteness of the $1/3$ qp). Due to the incompressible nature of the FQH liquid, there is a one-to-one correspondence between the qp states in the bulk and at the edge~\cite{wen95}. Hence an addition of a localized 1/3 qp in the puddle (a vortex) induces a propagating 1/3 qp (a soliton) at the edge. 
Therefore each addition of $5\phi_0$ flux introduces an additional edge $1/3$ qp and enhancement in the edge transport.

Starting from this observation, we can derive the phase $\gamma_N$ of a qp encircling the area $S$
when {\it  the flux through the $2/5$ puddle $|B|s$ is precisely $5N\phi_0$ } with an integer $N$.
This phase 
consists of two independent contributions: {\bf 1) the AB phase} which  depends on the area $S$ enclosed by the path of the edge qp,  $ -\frac{q^\star}{\hslash} \oint_S\vec{A}\cdot d\vec{l}\!=\! -2\pi|\frac{q^\star}{e}|\frac{|B|S}{\phi_0}\!=\! -10\pi|\frac{q^\star}{e}|\frac{S}{s}N$,  and {\bf 2) the statistical phase}
$2\theta^\star N$, due to the 1/3 qp forming the $2/5$ puddle of area $s$. Namely,
\begin{align}
\gamma_N(q^\star,\theta^\star,S/s)
 &=-2\pi\left(5\left|\frac{q^\star}{e}\right|\frac{S}{s}+\frac{\theta^\star}{\pi}\right)N\nonumber\\
 &=-\frac{2\pi}{3}\left(5\frac{S}{s}-1\right)N
 \label{eq:gamma}
\end{align}
where we used $\theta^\star\!=\!\pi/3$ and $q^\star\!=\!-|e|/3$.
Note that these two phases must be treated on an equal footing (see Ref.~\cite{arovas84}). Also the importance of  
the discreteness of $1/3$ qp reflected in the floor function of the statistical phase
cannot be understated. 

Constructive interference would occur if $\gamma_N$ is near an integer multiple of $2\pi$ (or $S/s$ a rational fraction). However, nothing forces the area ratio to be a rational fraction. For any value of the ratio, there will be many nearby rational fractions. In general, larger fractions will be closer to the actual value of the ratio but those require large periods and they are more susceptible to dephasing. Therefore the value of the rational fraction that dominates the oscillation will be determined by the competition between 1) quantitatively how close is the rational fraction to the actual value of the ratio and 2) how simple the resulting period is.  For the case of an area ratio of \emph{roughly} 1.43 (the setup of Ref.~\cite{camino-prl,camino-T,zhou}), oscillations associated with the simple ratio $7/5=1.4$ have the smallest possible period of $5\phi_0$ while higher ratios such as $10/7=1.428$ have a very large period of $105\phi_0$. Hence, this argument predicts that if oscillations occur in the given setup, the period will be $5\phi_0$. %Also it is clear from Eq.~\eqref{eq:gamma}, that it is likely for $5\phi_0$ oscillation to occur when the area ratio is $S/s\sim 7/5, 10/5, 13/5$. 
Notice also that proper accounting of fractional statistics is essential for this non-trivial superperiod for $1/3$ qp, that is different from the AB period $\Delta|B|\!=\!3\phi_0/S$, to emerge.

Before ending this discussion, we should comment on the interference at intermediate flux values $N\!<\!|B|s/5\phi_0\!<\!(N\!+\!1)$. Since the minimum amount of flux that can add a $1/3$ qp to the inner $2/5$ puddle is $5\phi_0$, the flux through the puddle \emph{cannot change} in this interval. Hence, in order to accomodate the increase in the $B$ field, the inner puddle area $s(B)$ must \emph{shrink} and change the electrostatic profile of the system until another $5\phi_0$ may be added to the puddle where it will return to its original size. Hence, this interval includes regions of destructive interference, but the behavior also depends on microscopic quantities.

We now calculate the tunneling conductance to the lowest order in the tunneling amplitudes using the effective theory for $1/3$ edge states in terms of a chiral Luttinger liquid with the Luttinger parameter $\nu\!=\!1/3$ ~\cite{wen90,wen95}. The left/right moving edge qp propagator is given by  
$\langle\psi_{L/R}^\dagger(x,t)\psi_{L/R}(0,0)\rangle
\!=\!\left\{\frac{\pi T\tau_0}{\sin(\pi T\tau_0+i\pi T(vt\pm x))}\right\}^\nu$,
where $\tau_0$ is a short distance cutoff and $v$ is the edge velocity.  Due to the branch cut in the propagator, the edge qp obeys twisted boundary conditions which determines the zero mode of the associated chiral boson. From an explicit derivation of the edge theory from the boundary terms of the Chern-Simons theory~\cite{zhang89,lopez91} for the bulk the twisted boundary condition for the edge qp can be directly related to the quantum mechanical consideration of the phase accumulation for an extra qp as
$\psi_{L/R}^\dagger(x\!+\!2\pi R)\!=\!e^{i\gamma_N}\psi_{L/R}^\dagger(x)$. Denoting the qp creation operators at the two tunneling points  $x_1$ and $x_2$ by $\psi_{R/L,i}^\dagger\!=\!\psi_{R/L}^\dagger(x_i,t)$ for $i=1,2$,  this twisted boundary condition can be represented in the tunneling Hamiltonian in the following manner
$H_t\!=\!\Gamma_1e^{-i\omega_Jt}\psi_{R,1}^\dagger\psi_{L,1}\!+\!
e^{i\gamma}\Gamma_2^*e^{i\omega_Jt}
\psi_{L,2}^\dagger\psi_{R,2}+h.c.$, where $\Gamma_1$ and $\Gamma_2$ are tunneling amplitudes.
Here, the Josephson frequency $\omega_J\equiv q^\star V/\hslash$ is introduced to impose the Hall voltage drop. A similar approach was taken for an interferometer without the central puddle  in Ref.~\cite{chamon-freed-kivelson-sondhi-wen97} and  our setup is a generalization of the setup studied by \textcite{chamon-freed-kivelson-sondhi-wen97}.

Now the tunneling conductance can be calculated perturbatively to the lowest order in the tunneling amplitudes to be
$G(\omega_J,v/R,T)\!=\!\bar{G}(\omega_J,T)\!+\!\cos\gamma(N)\; \delta G(\omega_J,v/R,T)$,
 for an integer $N$.
Only the (second) interference term depends on the edge velocity and on the separation between two point contacts $\pi R$. 
Considering the case  $\Gamma_1=\Gamma_2\equiv\Gamma$ for the sake of simplicity,  the tunneling conductance $\bar{G}$ is 
\begin{widetext}
\begin{equation}
\bar{G}(\omega_J,T) 
=\frac{{e^*}^2}{h}|\Gamma|^2(\pi T)^{2\nu-1} B \left(\nu-i\frac{\omega_J}{2\pi T}, \nu-i\frac{\omega_J}{2\pi T}\right)
\times\left[\frac{1}{2T}\cosh\left(\frac{\omega_J}{2T}\right)+2 {\rm Im}\; \psi\left(\nu-i\frac{\omega_J}{2\pi T}\right)\sinh\left(\frac{\omega_J}{2T}\right)\right]
\end{equation}
where $\psi(z)$ is the digamma function, and $B(z,z')$  the Euler Beta function.  The oscillation amplitude $\delta G$ is \vspace{-2mm}
\begin{align}
\delta G(\omega_J,v/R,T) &=4 \frac{{e^*}^2}{h}|\Gamma|^2(\pi T)^{2\nu-1} B \left(\nu-i\frac{\omega_J}{2\pi T}, \nu-i\frac{\omega_J}{2\pi T}\right)\sinh\left(\frac{\omega_J}{2T}\right)\nonumber\\
&\qquad\qquad\times\left[\left\{\frac{1}{2T}\coth\left(\frac{\omega_J}{2T}\right)+2 {\rm Im} \;\psi\left(\nu-i\frac{\omega_J}{2\pi T}\right)\right\}H_\nu(\omega_J,\pi R/v,T)+\frac{\partial}{\partial\omega_J}H_\nu(\omega_J,\pi R/v,T)\right]\label{eq:delG}
\end{align}
\end{widetext}
where the function $H_\nu$ is given in terms of the hypergeometric function $F$ as the following~\cite{chamon-freed-kivelson-sondhi-wen97}
\begin{align}
H_\nu&(\omega,x,T)=2\pi{\Gamma(2\nu) \over \Gamma(\nu)} {e^{-2\nu\pi T|x|} \over
\sinh {\omega \over 2 T}}\\
&\times\ {\rm Im}
\bigg\{
{e^{i\omega |x|}\ F(\nu,\nu\!-\!i{\omega \over 2\pi T};
1\!-\!i{\omega \over 2\pi T};e^{-4\pi T|x|})
\over
\Gamma(\nu\!+\!i{\omega \over 2\pi T})\ \Gamma(1\!-\!i{\omega \over 2\pi T})}
\bigg\}\nonumber
\end{align}

Since the temperature dependence of the tunneling amplitude will be a higher order effect, we remove the tunneling amplitude dependence by looking at the 
ratio $\delta G(T)/\delta G(T=11mK)$ and compare the calculated result with the experimental data reported in Ref.~\cite{camino-T}.  Taking the value of outer radius from Ref.~\cite{camino-T} to be $R=685nm$ and using only the edge qp velocity as a fitting parameter, 
 we found surprisingly good agreement with the data for $V=7.42\mu V$ and $v=1.41\times10^7m/s$. 
The fit is better with  the Hall voltage $V=7.42\mu V$ than with the Hall voltage $V=7.5\mu V$ which is the estimated value of Hall voltage in Ref.~\cite{camino-T}.

\begin{figure}[b]
\psfrag{G}{$\delta G(T)/\delta G(T=11mK)$}
\psfrag{T}{$T(mK)$}
\includegraphics[width=0.4\textwidth]{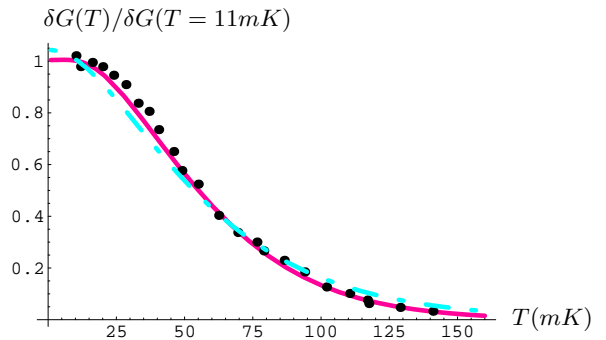}
\caption{The comparison between $\delta G(T)/\delta G(T=11mK)$ calculated from Eq.~\eqref{eq:delG} for $\nu\!=\!1/3$ and the data from Ref.~\cite{camino-T}. The dots are the experimental data and the (red) solid line is the calculated curve at $V=7.42\mu V$,  $v=1.41\times10^7m/s$ and the dashed (green) line is the calculated curve at $V=7.5\mu V$, $v=0.94\times10^7m/s$.}
\label{fig:cond}
\end{figure}

The fact that the curve calculated from our model fits  well with the data supports a number of aspects of our model. 
Firstly we assumed that the oscillation is a result of interference between qp's tunneling at two point contacts that are distance $\pi R$ apart. This assumption introduces an additional energy scale $v/\pi R$ to the oscillatory term. The lowest order perturbation theory captures the cross over associated with this scale,
allowing us to produce a full cross over curve. Hence an experimental observation of this crossover is a strong indication of  the interference between coherent qp's.  Secondly, the contiguous interference trajectory shows that the qp's maintain phase coherence not only while they propagate along the edge (in accordance with the hydrodynamic picture of edge states~\cite{wen90,wen95}) but also when they tunnel at two point contacts, and the decoherence only comes from thermal broadening. 
Thirdly, since the qp tunneling is a relevant perturbation in the RG sense, the fact that the lowest order perturbation theory nicely describes the experiment implies that the temperature was high compared to the scale determined by the tunneling matrix element. 
Finally,  the fact that  $\nu\!=\!1/3$ was used for the curve of Fig.~\ref{fig:cond} supports the assumption that transport is carried by $1/3$ qp's. 

Now we should address the question of the connection between the observed superperiodic AB oscillations and the  fractional statistics. Based on our detailed analysis,  the superperiodic AB effect observed in Refs.~\cite{camino-T,camino-prl,zhou} is likely to be a consequence of fractional statistics for the following reasons.
First of all, the conductance oscillation whose amplitude indicates Luttinger parameter $1/3$ showing the periodicity of $5\phi_0/s$, sensitive to the flux through the {\it island}, is a nontrivial effect yet it is consistent with our picture.  This superperiod coinsides with what one would expect from the combined effect of the AB phase and the statistical phase, assuming the ratio of areas is what was estimated in experiment. 
Secondly, the fact that there is a crossover in $\delta G$ as a function of $T$  implies that there is another scale in the problem. Given that the tunneling is weak enough that the system stays in the weak tunneling  limit, the only scale that can possibly enter is $R/v$. If indeed $R/v$ is setting this crossover scale, that is a strong indication that the conductance oscillation is due to interference. 

The model study hinges on the following key theoretical ingredients which are closely tied to one another: the fractional statistics obeyed by identical $1/3$ qp's, the incompressibility of the $2/5$ puddle and the hierarchical construction of $2/5$ puddle out of $1/3$ qp's. If our model is correctly describing the inner workings of the observed temperature dependent superperiod oscillations, this observation will serve as  the first direct confirmation of these fundamental theoretical ideas. However, for more definite confirmations, the following predictions should be tested. First, we expect the oscillation period to depend on the ratio $S/s$. Specifically, when $S/s$ is too far from a  commensurate values, oscillation will be absent. This will be a clear signature of presence of two periodicities, the AB period and the statistical period.
We also expect the 
crossover scale in the $\delta G-T$ curve to decrease with an increase in $R$ (see Fig.~\ref{fig:cond}). In addition, the voltage dependence can be compared with Eq.~\eqref{eq:delG}.  
\begin{figure}[t]
\psfrag{G}{$\delta G(T)/\delta G(T=11mK)$}
\psfrag{T}{$T(mK)$}
\includegraphics[width=0.4\textwidth]{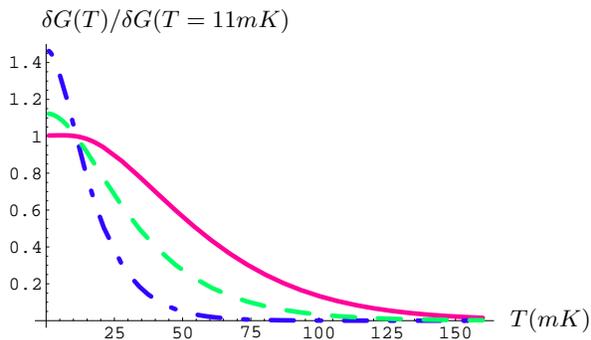}
\caption{The tunneling conductance oscillation amplitude   
 $\delta G(T)/\delta G(T=11mK)$ for different outer edge sizes at same Hall voltage of $V=7.42\mu V$. The solid (red) curve is for $R=685nm$, the dashed (green) curve is for $R=774nm$ and the dash-dotted (blue) curve is for $R=1370nm$. }
\label{fiq:cond}
\end{figure}
Finally, the oscillation will disappear in a dirtier sample since impurities will spoil the incompressibility of $2/5$ puddle by trapping qp's in the surrounding region.

A few remarks are in order before we close the paper. 
(1) Here we used the hierarchical picture to describe $2/5$ state. How to understand the experiment from the alternate description of $2/5$ state in terms of composite fermions without invoking hierarchy and condensation of $1/3$ qps~\cite{jain89,lopez91}. has been looked into recently~\cite{jain06} but the outcome was somewhat inconclusive and further investigation is needed. (2) We also assumed that there is {\it no qp tunneling between outer edge and the inner puddle} that contributes to the conductance oscillation. Indeed such tunneling process can also result in a series of periodic {\it  peaks} in the Hall conductance via a form of Coulomb blockade effect due to the finite size of the $2/5$ island in a manner similar to the case discussed in Ref.~\cite{kim03-prl}. However, 
the following points support our assumption. First, the observed rather smooth oscillation appears more like an interference effect. Second, while tunneling between outer and inner edge cost finite charging energy, tunneling between outer edges at point contacts is a relevant perturbation in RG sense. Hence it is more likely point contact tunneling between two outer edges is a dominant channel at low energy. Nevertheless, the Coulomb blockade possibility cannot be completely ruled out without more detailed analysis. Since Coulomb blockade scenario will be independent of outer edge radius, experiments with different outer edge radius will serve as a direct test. 
Clearly the anyonic nature of the FQH qp's are just becoming experimentally accessible.
While the only interferometer reported to have operated under the FQH conditions has been the focus of this paper, 
it will be vital to analyze observations from other geometries such as one in Ref.~\cite{ji03} in the future.\\
\noindent
{\bf Acknowledgments}:
 I am deeply indebted to S.~Kivelson for many enlightening discussions.  I thank V.~Goldman for  discussions and for the data. I am grateful to E.~Fradkin, C.~Nayak, B.~Halperin, J.~Jain, M.~Freedman and W.~Kang for useful comments. I thank M.~Lawler and S.~Vishveshwara for discussions.
This work was supported by Stanford Institute for Theoretical Physics.


\begin{thebibliography}{25}
\expandafter\ifx\csname natexlab\endcsname\relax\def\natexlab#1{#1}\fi
\expandafter\ifx\csname bibnamefont\endcsname\relax
  \def\bibnamefont#1{#1}\fi
\expandafter\ifx\csname bibfnamefont\endcsname\relax
  \def\bibfnamefont#1{#1}\fi
\expandafter\ifx\csname citenamefont\endcsname\relax
  \def\citenamefont#1{#1}\fi
\expandafter\ifx\csname url\endcsname\relax
  \def\url#1{\texttt{#1}}\fi
\expandafter\ifx\csname urlprefix\endcsname\relax\def\urlprefix{URL }\fi
\providecommand{\bibinfo}[2]{#2}
\providecommand{\eprint}[2][]{\url{#2}}

\bibitem[{\citenamefont{Camino et~al.}(2005{\natexlab{a}})\citenamefont{Camino,
  Zhou, and Goldman}}]{camino-prl}
\bibinfo{author}{\bibfnamefont{F.~E.} \bibnamefont{Camino}},
  \bibinfo{author}{\bibfnamefont{W.}~\bibnamefont{Zhou}}, \bibnamefont{and}
  \bibinfo{author}{\bibfnamefont{V.~J.} \bibnamefont{Goldman}},
  \bibinfo{journal}{Phys. Rev. Lett.} \textbf{\bibinfo{volume}{95}},
  \bibinfo{pages}{246802} (\bibinfo{year}{2005}{\natexlab{a}}).

\bibitem[{\citenamefont{Camino et~al.}(2005{\natexlab{b}})\citenamefont{Camino,
  Xhou, and Goldman}}]{camino-T}
\bibinfo{author}{\bibfnamefont{F.~E.} \bibnamefont{Camino}},
  \bibinfo{author}{\bibfnamefont{W.}~\bibnamefont{Xhou}}, \bibnamefont{and}
  \bibinfo{author}{\bibfnamefont{V.~J.} \bibnamefont{Goldman}}
  (\bibinfo{year}{2005}{\natexlab{b}}), \bibinfo{note}{cond-mat/0510764}.

\bibitem[{\citenamefont{Zhou et~al.}(2005)\citenamefont{Zhou, Camino, and
  Goldman}}]{zhou}
\bibinfo{author}{\bibfnamefont{W.}~\bibnamefont{Zhou}},
  \bibinfo{author}{\bibfnamefont{F.~E.} \bibnamefont{Camino}},
  \bibnamefont{and} \bibinfo{author}{\bibfnamefont{V.~J.}
  \bibnamefont{Goldman}}, \bibinfo{journal}{PRB} \textbf{\bibinfo{volume}{73}},
  \bibinfo{pages}{245322} (\bibinfo{year}{2005}),
  \bibinfo{note}{cond-matt/0512329}.

\bibitem[{\citenamefont{Leinaas and Myerheim}(1977)}]{leinaas77}
\bibinfo{author}{\bibfnamefont{J.}~\bibnamefont{Leinaas}} \bibnamefont{and}
  \bibinfo{author}{\bibfnamefont{J.}~\bibnamefont{Myerheim}},
  \bibinfo{journal}{Nuovo Cimento Soc. Ital. Fis.}
  \textbf{\bibinfo{volume}{37B}}, \bibinfo{pages}{1} (\bibinfo{year}{1977}).

\bibitem[{\citenamefont{Wilczek}(1982)}]{wilczek82}
\bibinfo{author}{\bibfnamefont{F.}~\bibnamefont{Wilczek}},
  \bibinfo{journal}{Phys. Rev. Lett.} \textbf{\bibinfo{volume}{48}},
  \bibinfo{pages}{1144} (\bibinfo{year}{1982}).

\bibitem[{pfa()}]{pfaffian-short}
\bibinfo{howpublished}{G. Moore, and N. Read, Nucl. Phys. B {\bf 360},
  362(1991); C. Nayak and F. Wilczek, Nucl. Phys. B {\bf 479}, 529 (1996)}.

\bibitem[{\citenamefont{Laughlin}(1983)}]{laughlin83}
\bibinfo{author}{\bibfnamefont{R.~B.} \bibnamefont{Laughlin}},
  \bibinfo{journal}{Phys. Rev. Lett.} \textbf{\bibinfo{volume}{50}},
  \bibinfo{pages}{1395} (\bibinfo{year}{1983}).

\bibitem[{\citenamefont{Halperin}(1984)}]{halperin84}
\bibinfo{author}{\bibfnamefont{B.~I.} \bibnamefont{Halperin}},
  \bibinfo{journal}{Phys. Rev. Lett.} \textbf{\bibinfo{volume}{52}},
  \bibinfo{pages}{1583} (\bibinfo{year}{1984}).

\bibitem[{\citenamefont{Arovas et~al.}(1984)\citenamefont{Arovas, Schrieffer,
  and Wilczek}}]{arovas84}
\bibinfo{author}{\bibfnamefont{D.}~\bibnamefont{Arovas}},
  \bibinfo{author}{\bibfnamefont{J.~R.} \bibnamefont{Schrieffer}},
  \bibnamefont{and} \bibinfo{author}{\bibfnamefont{F.}~\bibnamefont{Wilczek}},
  \bibinfo{journal}{Phys. Rev. Lett.} \textbf{\bibinfo{volume}{53}},
  \bibinfo{pages}{722} (\bibinfo{year}{1984}).

\bibitem[{\citenamefont{Wen}(1995)}]{wen95}
\bibinfo{author}{\bibfnamefont{X.~G.} \bibnamefont{Wen}},
  \bibinfo{journal}{Advances in Physics} \textbf{\bibinfo{volume}{44}},
  \bibinfo{pages}{405} (\bibinfo{year}{1995}).

\bibitem[{\citenamefont{Kivelson}(1990)}]{kivelson90}
\bibinfo{author}{\bibfnamefont{S.}~\bibnamefont{Kivelson}},
  \bibinfo{journal}{Phys. Rev. Lett.} \textbf{\bibinfo{volume}{65}},
  \bibinfo{pages}{3369} (\bibinfo{year}{1990}).

\bibitem[{\citenamefont{Jain et~al.}(1993)\citenamefont{Jain, Kivelson, and
  Thouless}}]{jain93}
\bibinfo{author}{\bibfnamefont{J.~K.} \bibnamefont{Jain}},
  \bibinfo{author}{\bibfnamefont{S.~A.} \bibnamefont{Kivelson}},
  \bibnamefont{and} \bibinfo{author}{\bibfnamefont{D.~J.}
  \bibnamefont{Thouless}}, \bibinfo{journal}{Phys. Rev. Lett.}
  \textbf{\bibinfo{volume}{71}}, \bibinfo{pages}{3003} (\bibinfo{year}{1993}).

\bibitem[{\citenamefont{de~C.~Chamon et~al.}(1997)\citenamefont{de~C.~Chamon,
  Freed, Kivelson, Sondhi, and Wen}}]{chamon-freed-kivelson-sondhi-wen97}
\bibinfo{author}{\bibfnamefont{C.}~\bibnamefont{de~C.~Chamon}},
  \bibinfo{author}{\bibfnamefont{D.~E.} \bibnamefont{Freed}},
  \bibinfo{author}{\bibfnamefont{S.~A.} \bibnamefont{Kivelson}},
  \bibinfo{author}{\bibfnamefont{S.~L.} \bibnamefont{Sondhi}},
  \bibnamefont{and} \bibinfo{author}{\bibfnamefont{X.~G.} \bibnamefont{Wen}},
  \bibinfo{journal}{Phys. Rev. B} \textbf{\bibinfo{volume}{55}},
  \bibinfo{pages}{2331} (\bibinfo{year}{1997}).

\bibitem[{\citenamefont{Safi et~al.}(2001)\citenamefont{Safi, Devillard, and
  Martin}}]{safi01}
\bibinfo{author}{\bibfnamefont{I.}~\bibnamefont{Safi}},
  \bibinfo{author}{\bibfnamefont{P.}~\bibnamefont{Devillard}},
  \bibnamefont{and} \bibinfo{author}{\bibfnamefont{T.}~\bibnamefont{Martin}},
  \bibinfo{journal}{Phys. Rev. Lett.} \textbf{\bibinfo{volume}{86}},
  \bibinfo{pages}{4628} (\bibinfo{year}{2001}).

\bibitem[{\citenamefont{Vishveshwara}(2003)}]{vishveshwara03}
\bibinfo{author}{\bibfnamefont{S.}~\bibnamefont{Vishveshwara}},
  \bibinfo{journal}{Phys. Rev. Lett.} \textbf{\bibinfo{volume}{91}},
  \bibinfo{pages}{196803} (\bibinfo{year}{2003}).

\bibitem[{\citenamefont{Kim et~al.}(2005)\citenamefont{Kim, Lawler,
  Vishveshwara, and Fradkin}}]{kim05}
\bibinfo{author}{\bibfnamefont{E.-A.} \bibnamefont{Kim}},
  \bibinfo{author}{\bibfnamefont{M.}~\bibnamefont{Lawler}},
  \bibinfo{author}{\bibfnamefont{S.}~\bibnamefont{Vishveshwara}},
  \bibnamefont{and} \bibinfo{author}{\bibfnamefont{E.}~\bibnamefont{Fradkin}},
  \bibinfo{journal}{Phys. Rev. Lett.} \textbf{\bibinfo{volume}{95}},
  \bibinfo{pages}{176402} (\bibinfo{year}{2005}).

\bibitem[{\citenamefont{Law et~al.}(2006)\citenamefont{Law, Feldman,
 and Gefen}}]{law06}
\bibinfo{author}{\bibfnamefont{K.T.} \bibnamefont{Law}},
  \bibinfo{author}{\bibfnamefont{D.E.}~\bibnamefont{Feldman}},
   \bibnamefont{and} \bibinfo{author}{\bibfnamefont{Yuval}~\bibnamefont{Gefen}},
  \bibinfo{journal}{Phys. Rev. B} \textbf{\bibinfo{volume}{74}},
  \bibinfo{pages}{045319} (\bibinfo{year}{2006}).
  
\bibitem[{\citenamefont{Tsui et~al.}(1982)\citenamefont{Tsui, St\"ormer, and
  Gossard}}]{tsui82}
\bibinfo{author}{\bibfnamefont{D.~C.} \bibnamefont{Tsui}},
  \bibinfo{author}{\bibfnamefont{H.~L.} \bibnamefont{St\"ormer}},
  \bibnamefont{and} \bibinfo{author}{\bibfnamefont{A.~C.}
  \bibnamefont{Gossard}}, \bibinfo{journal}{Phys. Rev. Lett.}
  \textbf{\bibinfo{volume}{48}}, \bibinfo{pages}{1559} (\bibinfo{year}{1982}).

\bibitem[{\citenamefont{Haldane}(1983)}]{haldane83}
\bibinfo{author}{\bibfnamefont{F.~D.~M.} \bibnamefont{Haldane}},
  \bibinfo{journal}{Phys. Rev. Lett.} \textbf{\bibinfo{volume}{51}},
  \bibinfo{pages}{605} (\bibinfo{year}{1983}).

\bibitem[{\citenamefont{Jain}(1989)}]{jain89}
\bibinfo{author}{\bibfnamefont{J.~K.} \bibnamefont{Jain}},
  \bibinfo{journal}{Phys. Rev. Lett.} \textbf{\bibinfo{volume}{63}},
  \bibinfo{pages}{199} (\bibinfo{year}{1989}).

\bibitem[{\citenamefont{Zhang et~al.}(1989)\citenamefont{Zhang, Hansson, and
  Kivelson}}]{zhang89}
\bibinfo{author}{\bibfnamefont{S.~C.} \bibnamefont{Zhang}},
  \bibinfo{author}{\bibfnamefont{T.~H.} \bibnamefont{Hansson}},
  \bibnamefont{and} \bibinfo{author}{\bibfnamefont{S.}~\bibnamefont{Kivelson}},
  \bibinfo{journal}{Phys. Rev. Lett.} \textbf{\bibinfo{volume}{62}},
  \bibinfo{pages}{82} (\bibinfo{year}{1989}).

\bibitem[{\citenamefont{L\'opez and Fradkin}(1991)}]{lopez91}
\bibinfo{author}{\bibfnamefont{A.}~\bibnamefont{L\'opez}} \bibnamefont{and}
  \bibinfo{author}{\bibfnamefont{E.}~\bibnamefont{Fradkin}},
  \bibinfo{journal}{Phys. Rev. B} \textbf{\bibinfo{volume}{44}},
  \bibinfo{pages}{5246} (\bibinfo{year}{1991}).


\bibitem[{\citenamefont{Wen}(1990)}]{wen90}
\bibinfo{author}{\bibfnamefont{X.~G.} \bibnamefont{Wen}},
  \bibinfo{journal}{Phys. Rev. B} \textbf{\bibinfo{volume}{41}},
  \bibinfo{pages}{12838} (\bibinfo{year}{1990}).

\bibitem[{\citenamefont{Jain and Shi}(2006)}]{jain06}
\bibinfo{author}{\bibfnamefont{J.~K.} \bibnamefont{Jain}} \bibnamefont{and}
  \bibinfo{author}{\bibfnamefont{C.}~\bibnamefont{Shi}}, \bibinfo{journal}{Phys. Rev. Lett}
  \textbf{\bibinfo{volume}{96}}, \bibinfo{pages}{136802}
  (\bibinfo{year}{2006}), \bibinfo{note}{cond-mat/0603535}.

\bibitem[{\citenamefont{Kim and Fradkin}(2003)}]{kim03-prl}
\bibinfo{author}{\bibfnamefont{E.-A.} \bibnamefont{Kim}} \bibnamefont{and}
  \bibinfo{author}{\bibfnamefont{E.}~\bibnamefont{Fradkin}},
  \bibinfo{journal}{Phys. Rev. Lett.} \textbf{\bibinfo{volume}{91}},
  \bibinfo{pages}{156801} (\bibinfo{year}{2003}).

\bibitem[{\citenamefont{Ji et~al.}(2003)\citenamefont{Ji, Chung, Sprinzak, Heiblum, Mahalu, and Shtrikman}}]{ji03}
\bibinfo{author}{\bibfnamefont{Y.} \bibnamefont{Ji}},
  \bibinfo{author}{\bibfnamefont{Y.} \bibnamefont{Chung}},
   \bibinfo{author}{\bibfnamefont{D.} \bibnamefont{Sprinzak}},
    \bibinfo{author}{\bibfnamefont{M.} \bibnamefont{Heiblum}},
     \bibinfo{author}{\bibfnamefont{D.} \bibnamefont{Mahalu}},
  \bibnamefont{and} \bibinfo{author}{\bibfnamefont{H.}~\bibnamefont{Shtrikman}},
  \bibinfo{journal}{Nature} \textbf{\bibinfo{volume}{422}},
  \bibinfo{pages}{415} (\bibinfo{year}{2003}).


\end{thebibliography}
\end{document}